# Local dielectric spectroscopy of near-surface glassy polymer dynamics

P. S. Crider, M. R. Majewski, Jingyun Zhang, H. Oukris and N. E. Israeloff

Dept. of Physics

Northeastern University

Boston, MA 02115

A non-contact scanning-probe-microscopy method was used to probe local near-surface dielectric susceptibility and dielectric relaxation in poly-vinyl-acetate (PVAc) near the glass transition. Dielectric spectra were measured from $10^{-4}$ Hz to $10^{2}$ Hz as a function of temperature. The measurements probed a 20 nm thick layer below the free-surface of a bulk film. A small (4 K) reduction in glass transition temperature and moderate narrowing of the distribution of relaxation times was found. In contrast to results for ultra-thin-films confined on or between metallic electrodes, no reduction in the dielectric strength was found, inconsistent with the immobilization of slower modes.

## Introduction

Recently, there have been a number of experiments which demonstrate a significant change in the glass transition temperature, $T_g$, of ultra-thin polymer films[1-7]. The effects are most dramatic, in the form of $T_g$ reductions, in free-standing high molecular weight polystyrene films[2,6] for films < 80 nm. In other cases where the films are constrained by substrates, the shifts in $T_g$ can be significantly smaller or even positive[1] if the substrate-polymer interactions are strongly attractive. But controversies remain because of differences seen with different measurement techniques. In some materials such as polyvinyl acetate (PVAc) a $T_g$ reduction is small[3,5] or undetectable[8] depending on the technique and thickness, though significant changes in mechanical moduli were observed[8].

The mechanisms for the shifts in $T_g$ are still debated, but intuitively one could expect higher mobility of polymer chains or monomers at a free-surface where geometric constraints are reduced. It is also well known that glassy dynamics are dynamically heterogeneous[9] and cooperative over some length scale. In PVAc, the heterogeneities were found to be 3 nm in diameter on average[10], though it was unclear if these regions or compact or string-like[11]. Thus a higher mobility might be expected to extend several nanometers below a free surface. The polymer chain length or radius of gyration, which can be much larger, can be an important length scale[3,12-14] especially in the high molecular weight cases[6], which is suggestive of *chain* confinement effects[13]. It is less clear how mobility can be increased near solid interfaces[12], however the immobilization of slower modes by confinement or interfacial pinning, leading to higher average mobility has been suggested[5].

A well-established technique for studying glassy dynamics in bulk, conventional dielectric susceptibility can characterize both $T_g$ shifts as well as changes in the relaxation time distribution in ultra-thin-films[3,5,7]. But this approach may have complications due to the interfacial effects of the electrodes, such as a dramatically reduced dielectric strength. In a double-electrode parallel-plate geometry $T_g$ was reduced at least 7 K in 20 nm thick PVAc films, the dielectric strength was reduced by two thirds, and the spectrum of relaxation times appeared to be broadened[3]. In contrast, a single-electrode measurement of PVAc films with one free-surface[5], the reductions of dielectric strength and $T_g$ were roughly half as much in 20 nm thick films, and the relaxation-time distribution was asymmetrically narrowed. This suggested that the small apparent reductions in $T_g$ and dielectric strength arose solely from a freezing out of the slower portion of the distribution, perhaps by interfacial pinning[5]. Here we report results of experiments in which a non-contact scanning probe microscopy technique is used to measure local frequency-dependent dielectric susceptibility in PVAc films. The technique, described in more detail elsewhere[15], probes to a depth of 20 nanometers below a free-surface of *bulk* films, and captures with high resolution both $T_g$ shifts and changes in the dynamics. In PVAc, a small reduction in $T_g$, a narrowing of the spectrum of relaxation times, but no reduction in dielectric strength were observed.

## Experiments

Electric force microscopy (EFM) is a form of scanning probe microscopy (SPM) in which non-contact electrostatic forces are measured. It has been used to image localized charges[16] on surfaces, dielectric constant variations and potentiometry[17]. In EFM a conducting cantilever is used so that DC and AC bias voltages can be applied to the tip. We use an ultra-high-vacuum SPM with a variable temperature stage (RHK UHV 350). See figure 1. A non-contact frequency-modulation (FM) imaging mode is used. In this mode, the cantilever is oscillated at its resonance-frequency, $f_{res}$ ~ 70 kHz, using a Nanosurf EasyPLL phase-locked-loop. Resonance frequency shifts, $\delta f_{res}$, due to tip-sample interaction forces are detected with very high resolution using a Nanosurf EasyPLL detector, and this is used as a feedback parameter for controlling the tip-sample distance, z. Temperature was measured with a small thermocouple clamped to the sample surface with a sapphire washer. The sample substrate was radiatively heated from below. The samples are thin films (thickness ~ 1 μm) of PVAc (Sigma) ($M_w$ = 167,000), which have been prepared by spinning from a toluene solution and annealed near $T_g$ = 308 K for 24 hours in vacuum. The substrates are glass with an Au coating which forms the ground electrode. Identically prepared samples were capped with sputtered electrodes of AuPd, and these capacitors were studied by conventional dielectric susceptibility with a Novovcontrol dielectometer for comparison. Typical bulk dielectric susceptibility behavior was found, as described below.

With a bias voltage applied to the EFM tip, the resulting tip charge will produce an image charge in the sample which is a function of the dielectric susceptibility, thus giving an attractive force which increases as the tip approaches the sample. The resulting force derivative, dF/dz, reduces the effective spring constant of the cantilever (nominal k ~ 2 N/m) and thereby reduces its resonance frequency by $\delta f_{res}$. In order to isolate the electrostatic forces from van der Waals forces we apply an oscillating tip bias, $V=V_0\sin(\omega t)$, at frequencies much lower than $f_{res}$. Since both tip and dielectric image charges are proportional to bias voltage, the electrostatic component of $\delta f_{res}$ is proportional to the square of the applied voltage, and thus gives a 2ω signal. The amplitude and phase of the 2ω signal, which is measured with a lock-in amplifier, can then be used to extract the local dielectric susceptibility. The $V_{2\omega}$ signal is given by[15]:

$$V_{2\omega} = \frac{A}{4}\frac{\delta k}{k} f_{res} = \frac{A}{4k}\frac{\partial F}{\partial z} f_{res} \qquad (1)$$

where A is a gain factor from the Nanosurf detector. The detector also introduces an instrumental phase-shift into $V_{2\omega}$ which is characterized on a metallic sample, and subtracted from the measurements. Height feedback introduces additional phase-shifts and thus is turned-off during the measurements. At room temperature the polymer susceptibility is essentially frequency-independent and real, and in this case we could compare the measured $V_{2\omega}$ vs. z to a simple model using[17] $dF/dz = ½V^2 d^2C/dz^2$, where C is the tip to ground capacitance. Both a finite-element model (FEM) with a realistic tip-shape and a sphere image-charge model give similar results for tip radius, R=28±5 nm and for the measurements that follow we determined that $z_0 = 15 \pm 3$ nm. The sample signal (force derivative) is most sensitive to the near surface response. The integrated response vs. depth can be calculated from derivatives of the electric field with respect to z in the sample, obtained from the FEM model, and used to define an effective probed volume, which was 20 ± 5 nm in depth and 40± 5 nm in diameter. More precisely, we find that half of the total integrated signal comes within a depth of 12 nm and about 2/3 comes from within 20 nm. Ultra-thin films studied previously by other techniques have been in this range.[8] By adjusting the resonance setpoint, $\delta f_{res}$, the $z_0$ is adjusted, and thereby the probed depth can be slightly increased by at most 5 nm.

The bulk complex dielectric susceptibility for PVAc has the form $\varepsilon = \varepsilon_\infty + \Delta\varepsilon$, where $\varepsilon_\infty$ is the high-frequency polarizability component. $\Delta\varepsilon$ is the dipolar component which freezes out at the glass transition. Above $T_g$ it produces the low-frequency rise to a plateau in the real part and a peak in the imaginary part as can be seen in figure 1 (a) for 322 K. In figure 1 (b) we show the real and imaginary components of $V_{2\omega}$, for 315 K. These curves are qualitatively very similar to the bulk ε' and ε" though they occur at significantly lower temperature. We find that the measured peak shapes and positions are independent of oscillation amplitude (< 8 nm) and bias voltage ($V < 1V_{rms}$), thus we are in a linear response regime.

In order to quantitatively understand these measurements, we must calculate the force derivative for a complex ε. We can use a partially-filled parallel-plate model, in which the tip is modeled as a circular plate of radius R held a distance, z, above a dielectric of thickness d, backed by a ground plane. The z makes small oscillations about an average position: $z = z_0 + \delta z \sin(\Omega t)$, where $\Omega = 2\pi f_{res}$. The tip bias, V, will produce an electric field in the vacuum space, $E_v = V/(z + d\varepsilon_0/\varepsilon)$. The dielectric surface will have a bound charge density $\sigma = \varepsilon_0 E_v(\varepsilon - \varepsilon_0)/\varepsilon$. The force derivative is dominated by the interaction between $E_v$ and σ, since other charges are far away. Only the fast component, $\varepsilon_\infty$, contributes to changes in σ and $E_v$ due to tip (z) oscillations, thus $\delta\sigma = \varepsilon_0 \delta E_v(\varepsilon_\infty - \varepsilon_0)/\varepsilon_\infty$ and $\delta E_v = E_v \delta z/(z + d\varepsilon_0/\varepsilon_\infty)$. The force derivative is thus:

$$\frac{dF}{dz} = \frac{1}{2}\pi R^2 [E_v \frac{\partial \sigma}{\partial z} + \sigma \frac{\partial E_v}{\partial z}] = \frac{1}{2} \frac{\pi \varepsilon_0 R^2 V^2}{[z_0 + d\frac{\varepsilon_0}{\varepsilon}]^2 [z_0 + d\frac{\varepsilon_0}{\varepsilon_\infty}]} [\frac{\varepsilon}{\varepsilon - \varepsilon_0} + \frac{\varepsilon_\infty}{\varepsilon_\infty - \varepsilon_0}] \qquad [2]$$

Using eqns. 1 and 2 we calculate the real and imaginary components of $V_{2\omega}$ with different ε inputs. We can achieve quantitative fits using $z_0$=12 nm and d = 40 nm, R=17 nm. As can be seen by comparing the solid lines in figure 1 (a) and (b) the calculated $V''_{2\omega}$ peak frequency is shifted 0.2 decades higher than that of ε". We see, however, that using the bulk susceptibility (322 K) data plugged into the model calculation, gives a broader imaginary peak and a more gradually decreasing real component. If we use instead a Havrilak-Negami[18] functional form, $\varepsilon = \varepsilon_\infty + iC\omega^{-2} + \Delta\varepsilon(1+(i\omega\tau)^\alpha)^{-\gamma}$, where C is conductivity, as an input to the model we can fit the data much better with exponents, γ =0.60 and α=0.94. These width and asymmetry exponents indicate a moderately narrower and more symmetric distribution as compared with γ = 0.52 and α = 0.83, needed to fit the bulk data. This behavior of α is distinctly opposite of that seen by Fukao et. al.[3] in which a broadening was seen in thin PVAc films. Another important result is that, unlike previous studies with electrode interfaces[3,5], no reduction in the dielectric strength, Δε, is required to obtain a quantitative fit.

The imaginary components of $V_{2\omega}$ measured for several temperatures near $T_g$, when normalized by peak value and peak frequency, can be collapsed quite well onto a single curve as shown in figure 2. This can be compared to the model calculation using the bulk ε data as an input. Clearly this does not fit the measurements well, i.e. the bulk normalized peak is broader than the local normalized peak. Since we have contributions from both near surface and bulk layers we consider a two-component model. We find we can fit (Fig. 2) the data best with an 85% contribution from a near-surface narrow distribution with α = 0.94 and γ = 0.56, and a 15% bulk contribution whose peak is shifted one decade lower in frequency. This model can account for the narrowing near the peak but less narrowing in the low frequency wing.

The frequency where tanδ = ε"/ε' peaks is slightly higher than the peak in ε", and is often used to identify an average relaxation time. Local tanδ $=V''_{2\omega}/V'_{2\omega}$ curves are shown in figure 3. As compared with bulk data, the local peaks appear to be shifted to considerably higher (1.3 decades) frequencies. A small part of this (0.2 decades) arises from the measurement technique, as the model calculation shows. In order to extend these measurements to much lower frequencies and temperatures, we have also carried out local dielectric relaxation measurements, i.e. measurements of the local polarization vs. time following application of a dc voltage. Measurement and imaging of local spontaneous polarization has been described in detail elsewhere[19]. In this instance we apply a 100 Hz, 1.0 volt rms bias plus a dc bias of 0.25 or 0.5 V. Near $T_g$ the dc bias locally polarizes the sample if applied for longer than the relaxation time. The polarization is quasi-static compared with both the tip and voltage oscillation periods, and produces an apparent dc offset voltage $V_p$. The polarization-induced surface charge produces an image on the tip, $Q'_p = C_{tip}V_P$,. The total tip charge is taneous hen $Q_{tip} = C_{tip}(V+V_P)$. The tip charging energy is $U = \frac{1}{2} Q_{tip}^2/C_{tip} = \frac{1}{2}C_{tip}(V+V_P)^2$ and the force derivative is $dF/dz = \frac{1}{2}(V+V_P)^2\, d^2C/dz^2$ After squaring, the cross term causes a 1ω signal to appear:

$$V_\omega = \frac{A}{2k}\frac{\partial^2 C}{\partial z^2}V_0 V_P f_{res} \qquad (2)$$

This signal, which is proportional to the polarization, is also measured with the lock-in amplifier. As shown in figure 4, we see $V_\omega$ vs. time at three temperatures, normalized to the initial value. All have fit reasonably well to a stretched exponential, $V_\omega/V_\omega(0) = A+B\exp((-t/\tau)^\beta$ . We Fourier transformed these curves to obtain frequency domain functions related[3] to the real and imaginary components of the susceptibility. We then can obtain the peak in tanδ, for direct comparison with the spectral data at higher frequencies.

In figure 5, we plot tanδ peak frequencies for $V_{2\omega}$ data, adjusted downward by 0.2 decades, also tanδ derived from local relaxation, as well as $1/2\pi\tau$ vs. temperature. We see that points derived from spectra and those derived from relaxation appear to lie on a single curve, and $1/2\pi\tau$ points nearly coincide with tanδ points. We also plot tanδ peaks for the bulk data. Local and bulk tanδ curves are fitted to a Vogel-Fulcher-Tammann (VFT)[20] law. The local curve is shifted to lower temperatures relative to the bulk by 3-5 K with a larger shift at higher frequency. If we use the typical definition of $T_g$ as the temperature when $\tau$ = 100 s, or peak frequency of 0.0016 Hz, we see that the fits indicate $T_g$ is shifted lower by 3.4±0.5 K for the local data. On the other hand these results can be interpreted as a speeding up of the average relaxation time, $\tau_\alpha$, by 1.1 decades as compared to bulk.

**Discussion and Conclusions**

In conclusion, using a non-contact dielectric spectroscopy technique, we have found higher mobility and a small $T_g$ reduction in PVAc within 20 nm of a free surface. We find no observable reduction in the dielectric strength and some narrowing of the peak in the imaginary component of the susceptibility. This narrowing is surprising, given the spread in contributions from various depths. However, we were able to show how a two-component model which includes a small bulk-like contribution can nicely account for the shape of the observed distribution. In previous work on dielectric susceptibility of ultra-thin films PVAc a similar reduction of $T_g$ and increase in $\tau_\alpha$ was found for films of thickness similar to our probed depth. A slightly smaller increase in $\tau_\alpha$ was found by Serghei et. al.[5] who used a single electrode, whereas by Fukao et. al.[3] found a slightly larger increase with two electrodes, about 1.0 decades at 333 K. In experiments with electrode interfaces, the dielectric strength, $\Delta\varepsilon$, was considerably suppressed in films near 20 nm thickness, by 30 % [5] with one electrode and by 65% with two electrodes[3]. In contrast, with no contact electrodes we find no measurable reduction in $\Delta\varepsilon$. This strongly suggests that the electrode interfaces are pinning or immobilizing the dipolar degrees of freedom. Serghei et. al.[5] argue that the immobilization is asymmetric, affecting primarily the low frequency side of the distribution. This reduction apparently accounts fully for the shift in average relaxation $\tau_\alpha$ and apparent $T_g$. Our shifts in $T_g$ cannot be explained in this way. We find a speeding up of the entire distribution and a moderate narrowing, primarily on the high-frequency side with no reduction in $\Delta\varepsilon$. The narrowing alone is insufficient to explain the shift in $\tau_\alpha$. This increase in mobility arises from, and must be explained by, the presence of the free surface, as it is in free-standing films. However, it remains to be explained why these effects in PVAc are so much less than in free-standing polystyrene films of similar molecular weight.

**Acknowledgements**

This work was supported by NSF grant DMR-0606090 and ACS-PRF 38113-AC7.

Fig 1. In (a) the bulk real and imaginary components of the susceptibility for PVAc at 322 K are plotted vs. frequency. In (b) the local real and imaginary components of $V_{2w}$ vs. frequency at 315 K for $\delta f = 40$ Hz, $V_{bias}$=0.6$V_{rms}$ are shown. Also plotted are the model outputs using bulk susceptibility measurements (322K) (solid lines) as inputs, or best fit susceptibility inputs (dashed lines). Inset: EFM tip sample configuration is shown. The dashed line delineates the effective volume probed.

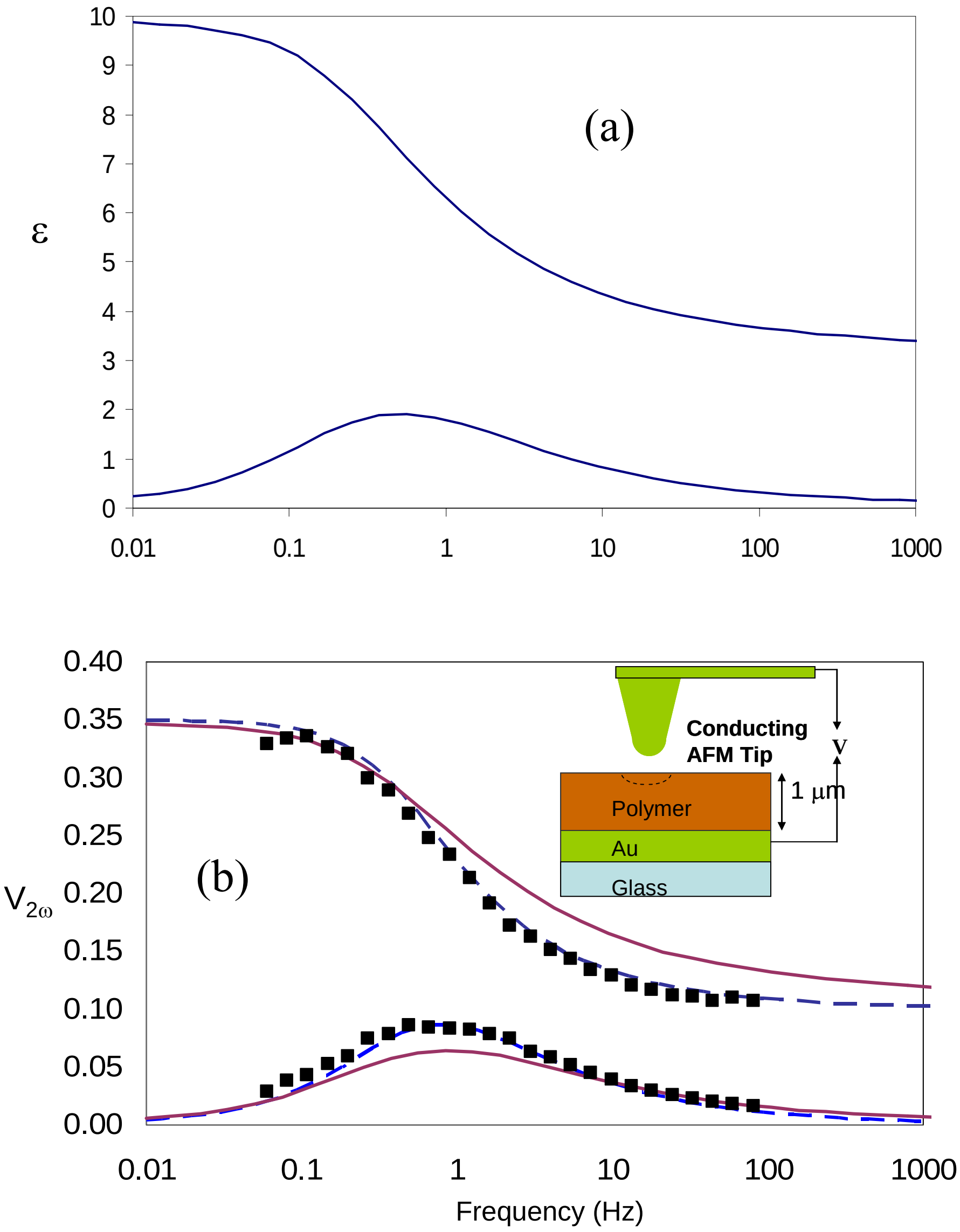

Figure 2. Peaks in the imaginary component of local response ($V_{2\omega}$), normalized to peak frequency and amplitude, are plotted vs. frequency for several temperatures for $\delta f = 55$ Hz, $V_{bias}$=1 $V_{rms}$. Model calculations (lines), using the bulk susceptibility data, and best fit susceptibility data, are shown for comparison.

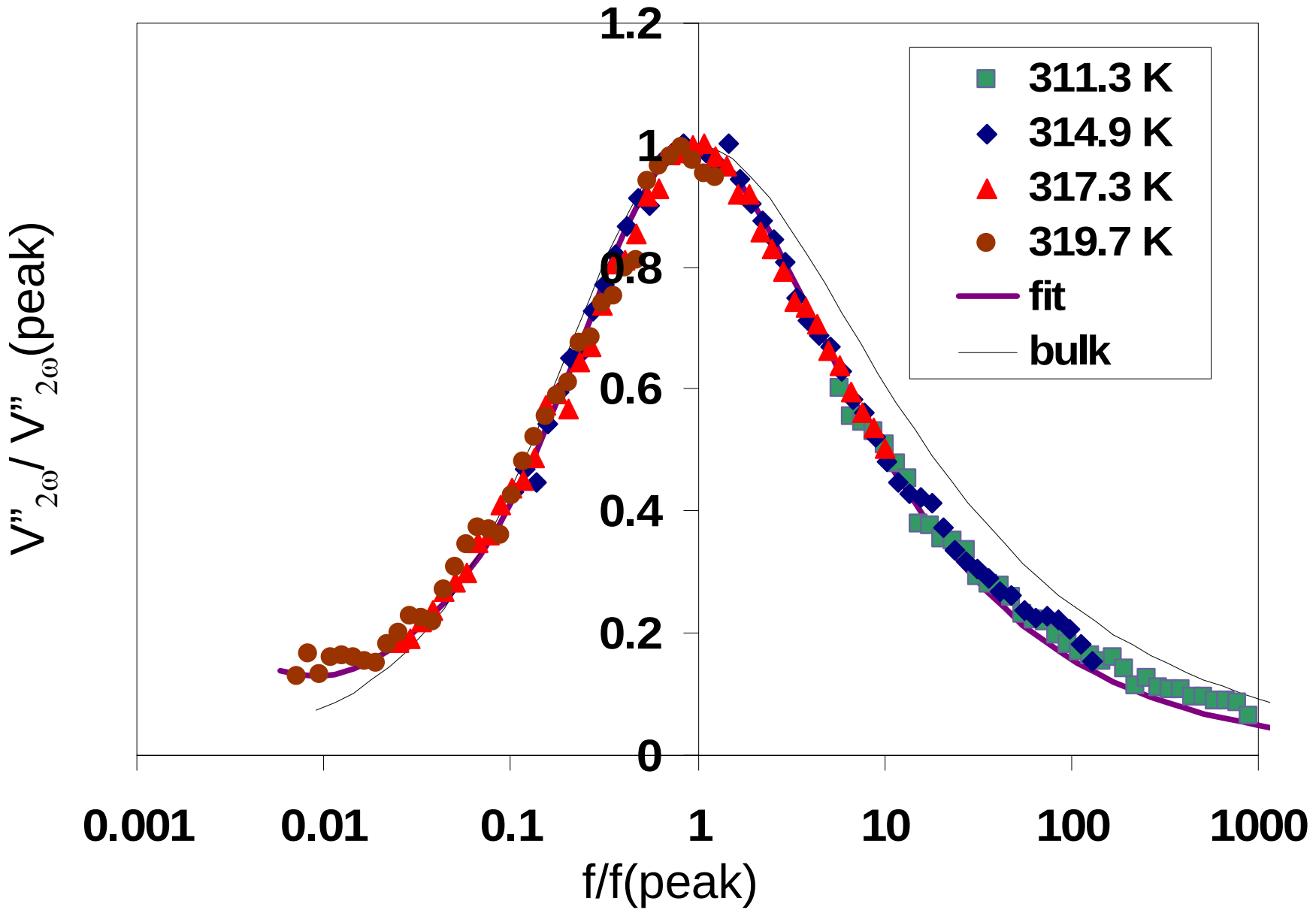

Figure 3. Local $\tan(\delta) = V''_{2\omega}/V'_{2\omega}$ vs. frequency for various temperatures for $\delta f$ = 55 Hz.

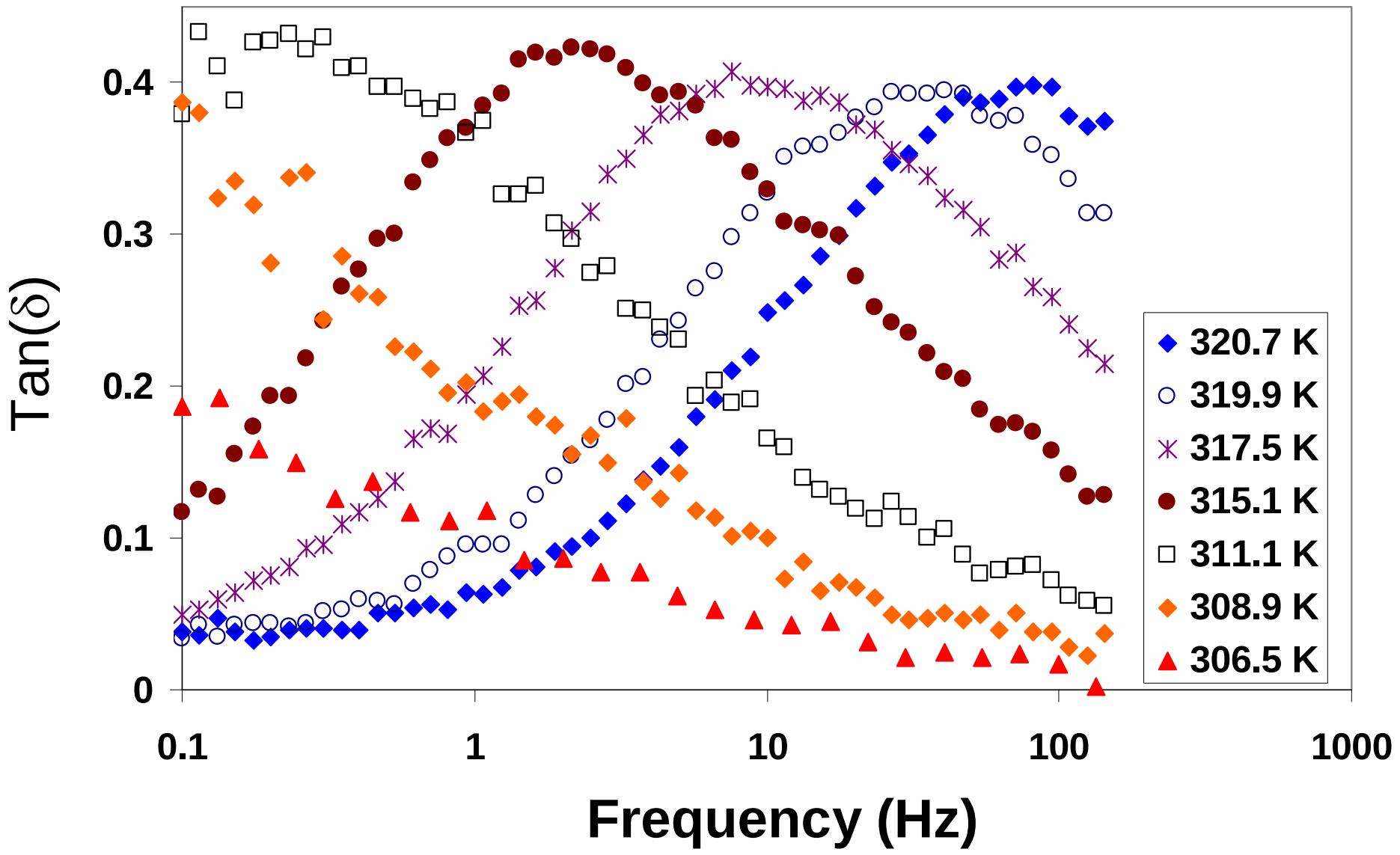

Figure 4. Local dielectric relaxation measurements at three temperatures. The $V_\omega$ was measured at zero bias, immediately following application of a 500 mV bias for an extended period, and is normalized by the initial value.

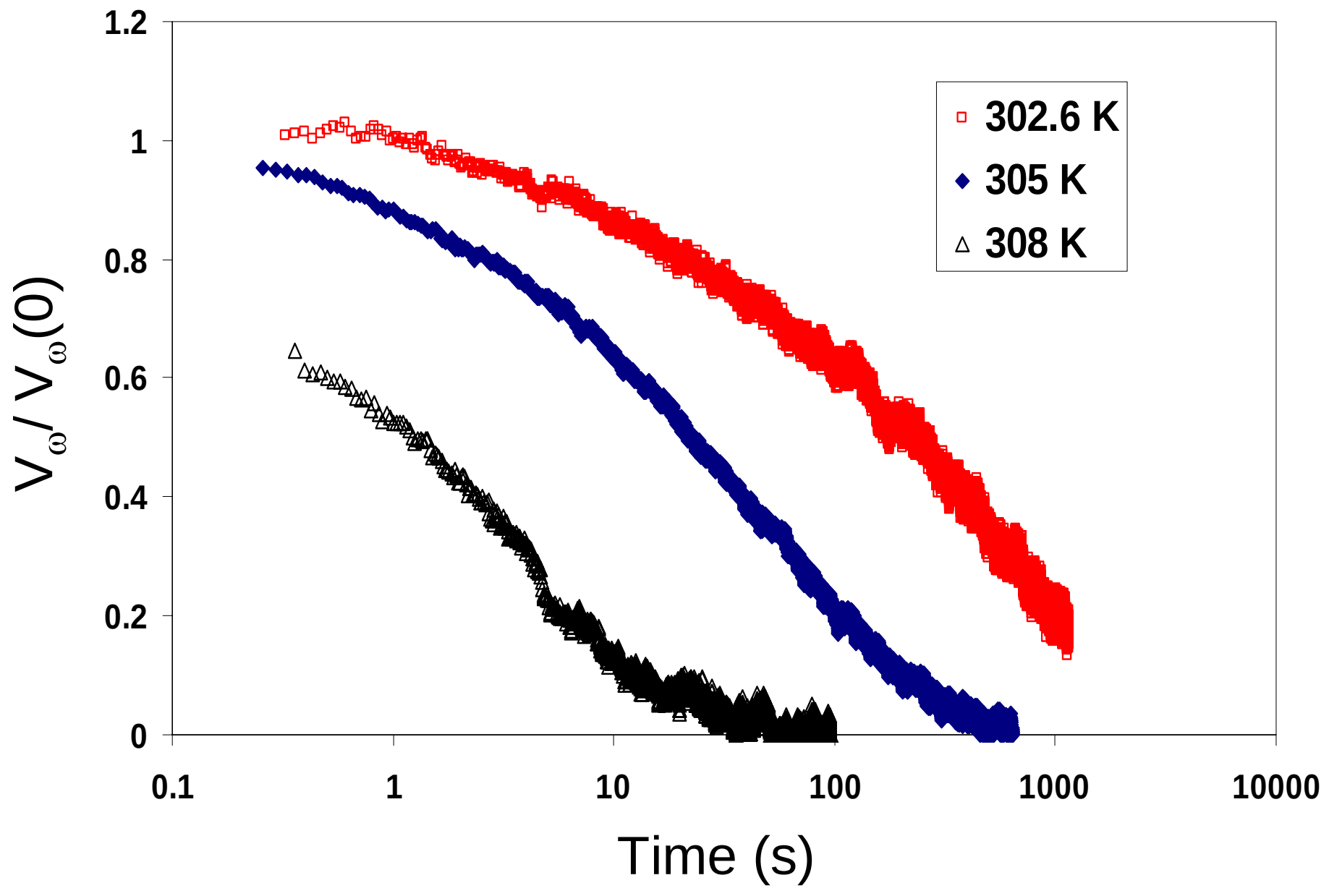

Figure 5. Dielectric susceptibility tan(δ) peak frequency vs. temperature for bulk and local measurements, δf=55 Hz. Points at lower temperature are obtained from Fourier transform of local time-domain dielectric relaxation measurements. Lines are fits to Vogel-Fulcher-Tamman law. Also plotted are 1/2πτ for τ obtained from stretched-exponential fits to local relaxation data.

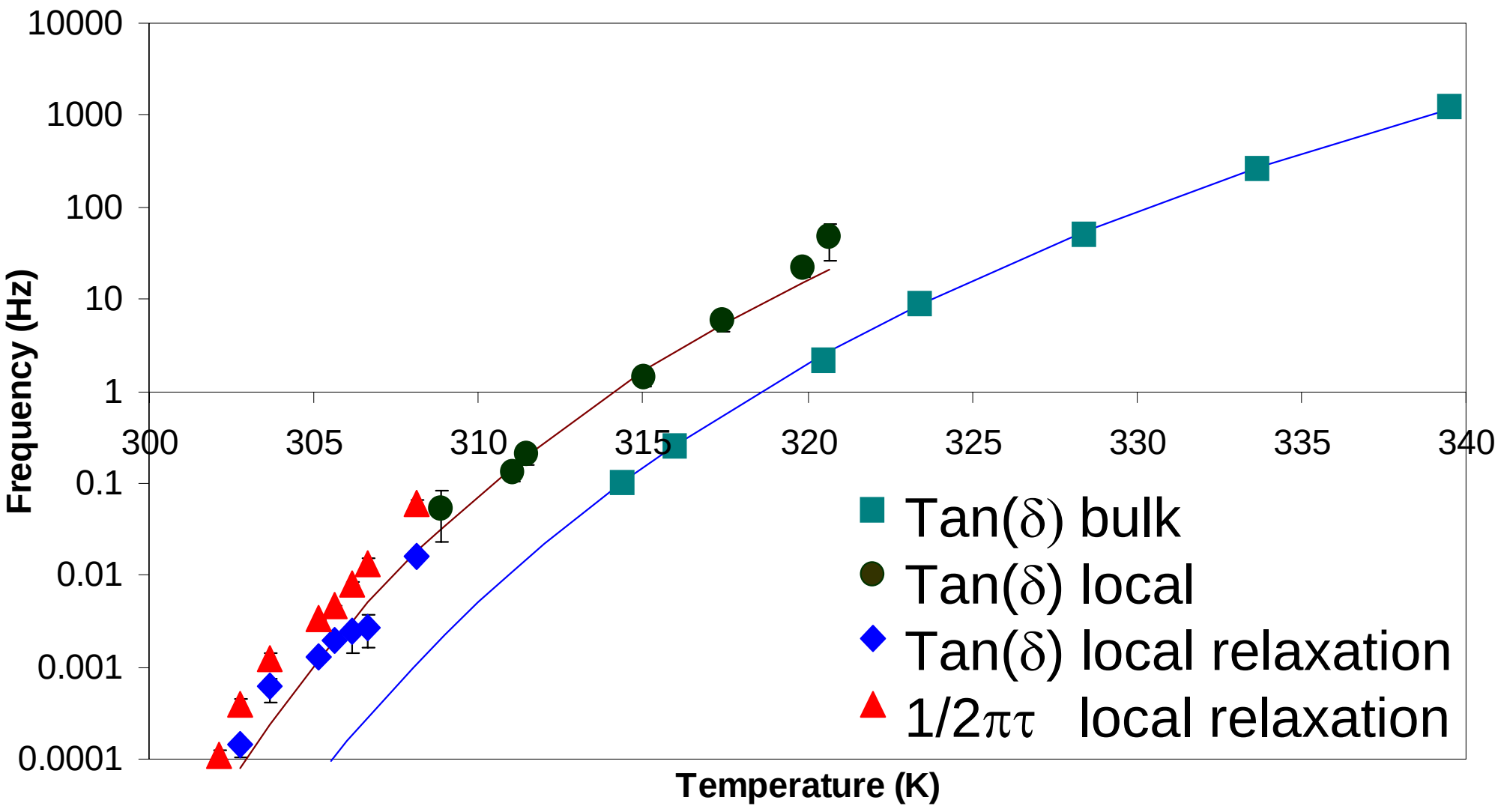